\documentclass[review]{elsarticle}
\usepackage{graphicx}
\usepackage{lineno,hyperref}
\usepackage{bm}
\usepackage{amsmath,amssymb}

\modulolinenumbers[5]

\journal{Physics Letters A}

\newcommand{\Ope}[1]{\mathcal{#1}}
\newcommand{\Bod}[1]{\mathbb{#1}}

\newtheorem{theorem}{Theorem}

\newtheorem{criterion}[theorem]{Criterion}

\begin{document}

\begin{frontmatter}

\title{Stability Boundaries and Sufficient Stability Conditions for Stably
Stratified, Monotonic Shear Flows}

\author{Makoto Hirota}
\address{Institute of Fluid Science, Tohoku University, Sendai, Miyagi 980-8577, Japan}

\author{Philip J. Morrison}
\address{%
Department of Physics and Institute for Fusion Studies, University of
Texas at Austin, Austin, Texas 78712, USA
}



\begin{abstract}
Linear stability of inviscid, parallel, and stably stratified shear flow is
 studied under the assumption of smooth strictly monotonic profiles of shear flow
 and density, so that the local Richardson number is positive
 everywhere. The marginally unstable modes are systematically found by
solving a one-parameter family of regular Sturm-Liouville problems,
 which can determine the stability
 boundaries more efficiently than solving the Taylor-Goldstein equation directly. By arguing for the non-existence of a marginally unstable mode, we derive
new sufficient conditions for stability, which generalize the
 Rayleigh-Fj\o rtoft criterion for unstratified shear flows.
\end{abstract}


\end{frontmatter}

\linenumbers

\section{Introduction}

Instabilities of stratified shear flows are important for understanding
not only geophysical phenomena but also fundamental mechanisms of flow
instabilities. The couplings among vortical waves and gravity waves,
which are induced by the ambient shear flow, give rise to various types
of instabilities named after Kelvin, Helmholtz, Holmboe, and Taylor (see
the review~\cite{Howard2}).  By replacing gravity by centrifugal force,
the analogous mechanism further triggers instabilities of swirling flows
and vortices.  The above instabilities have been described most clearly
by assuming staircase profiles of the ambient vorticity and density,
which enables one to interpret the instabilities in terms of the
couplings of a finite number of discrete
eigenmodes~\cite{Howard2}. Slightly-smoothed staircase profiles are
studied approximately by using matched asymptotic expansions~\cite{Balmforth}.

However, it is generally difficult
to predict and interpret the instabilities of continuous profiles
because the stable disturbances occupy a continuous
spectrum~\cite{Eliassen,Miles}. Since the corresponding eigenvalue
problem, called the Taylor-Goldstein equation, is non-self-adjoint and
singular, the accurate stability boundary is often hard to determine
both analytically and numerically.  The classical Miles-Howard
criterion~\cite{Miles,Howard} states that the stratified shear flow is stable
if the local Richardson number $J_R$ is greater than $1/4$ everywhere.
Although many theoretical investigations have been made
concerning various situations where $J_R\le1/4$ somewhere (see
\cite{Alexakis,Churilov,Rees} and references therein), 
a general stability criterion for smooth profiles does not appear to be in the literature.
However, a
generalized version of Rayleigh-Fj\o rtoft criterion is found when $0<J_R\le1/4$ everywhere in Ref.~\cite{Banerjee}.

 In this letter, we will present an efficient method for finding marginally unstable
modes in the case of $J_R>0$ everywhere. Following this method, we will
further obtain new sufficient conditions for stability.

\section{Search for Marginally Unstable Modes}

We consider the linear stability of parallel shear flow
$\bm{U}=(0,U(x))$ in an inviscid, incompressible fluid of
variable density $\rho(x)$ on a domain $(x,y)$ bounded by
two walls at $x=\pm L$, where the gravitational
acceleration $g$ acts in the $-x$ direction. By
introducing the stream function of the disturbance as
$\phi(x)e^{ik(y-Ct)}$ with a complex phase speed $C\in\Bod{C}$ and
a (real) wavenumber $k>0$, stability is governed by the
Taylor-Goldstein (TG) equation:
\begin{align}
 \phi''-k^2\phi+\frac{U''}{C-U}\phi+\frac{N_B^2}{(C-U)^2}\phi=0,
\phi(\pm L)=0,
\end{align}
where the prime ($'$) indicates the $x$ derivative and the Boussinesq
approximation [$\rho(x)=\rho_0+\delta\rho(x)$ where $\rho_0\gg|\delta\rho|$]
has been used;
$N_B=\sqrt{-g\rho'/\rho_0}$ is the Brunt-V\"ais\"al\"a (or buoyancy) frequency. If
this equation has a nontrivial solution for $C$ with a
positive imaginary part, ${\rm Im}\,C>0$, the shear flow is
spectrally unstable. In what follows, we assume a strictly increasing shear
flow and a stably stratified density;
\begin{align}
 U'>0\quad\mbox{and}\quad\rho'<0\quad\mbox{on }[-L,L],\label{assumption}
\end{align}
 so that the local Richardson
number $J_R=N_B^2/U'^2$ is positive everywhere.
(Strictly decreasing shear flows $U'<0$ can be treated similarly by
replacing $U$ by $-U$ in the TG equation. In fact, a flow $U$ is stable if and only if $-U$ is so.)

Stability boundaries are often studied by searching for marginally
unstable eigenmodes with $C=c+i0=\lim_{\epsilon\rightarrow
+0}(c+i\epsilon)$ (where $c={\rm Re}\,C\in\Bod{R}$), which were called
singular neutral modes by Miles~\cite{Miles} because the TG equation
becomes singular at the critical layer $x_c$ satisfying $c=U(x_c)$.
Miles proved that a singular neutral mode may exist only when
$J_R(x_c)\le1/4$ and only in the form of either $\phi\propto\phi_+$ or
$\phi\propto\phi_-$ (see the results VIII and IX of \cite{Miles}), where
\begin{align}
 \phi_\pm=(c+i0-U)^{1/2\pm\nu_c}\varphi_\pm(c,x),\label{SNM}
\end{align}
with the sign of $\nu_c=\sqrt{1/4-J_R(x_c)}$ delineating the two
linearly independent solutions obtained by the Frobenius
method. Specifically, the singularity of \eqref{SNM} has the following
branch cuts extending to infinity,
\begin{align}
& (c+i0-U)^{1/2\pm\nu_c}\nonumber\\
=&|c-U|^{1/2\pm\nu_c}\exp\left[i\pi\left(\frac{1}{2}\pm\nu_c\right)H(x_c-x)\right],
\end{align}
where $H(x)$ is the Heaviside function, and $\varphi_\pm$ are real
analytic functions satisfying $\varphi_\pm(c,x_c)\ne0$ as well as the
boundary condition $\varphi_\pm(c,-L)=\varphi_\pm(c,L)=0$.  Thus, the
existence of the domain $D_w=\{x\in[-L,L]:J_R(x)\le1/4\}$ is necessary
for instability, in that the singular neutral modes must have the
critical layers on it.  Since $0\le\nu_c<1/2$ for $x_c\in D_w$, we will
refer to the less singular mode $\phi_+$ as the ``small'' neutral mode,
and $\phi_-$ as the ``large'' neutral mode (following the terminology of
ideal MHD stability theory). Notice that these two types of neutral
modes degenerate to $(c+i0-U)^{1/2}\varphi(c,x)$ only when
$J_R(x_c)=1/4$ (which usually occurs when $x_c$ is an endpoint of
$D_w$).  For later use, we also define
$\check{D}_w=\{x\in[-L,L]:J_R(x)<1/4\}$ by omitting such the endpoints.

In order to find these singular neutral modes
efficiently, we first apply a transformation $\psi=(C-U)^{-1/2+\nu_c}\phi$
to the TG equation~\cite{Miles2} and obtain, for $C=c+i0$,
\begin{align}
 (P\psi')'-(k^2+Q)P\psi=0,\quad
 \psi(\pm L)=0,
\end{align}
where
\begin{align*}
 P(x,x_c)=&|U-c|^{1-2\nu_c}\exp\left[i\pi\left(1-2\nu_c\right)H(x_c-x)\right],\\
 Q(x,x_c)=&
\left(\frac{1}{2}+\nu_c\right)\frac{U''}{U-c}-U'^2\frac{J_R-J_R(x_c)}{(U-c)^2}.
\end{align*}
Because
$|P|^{-1},|P|$ and $|QP|$ are integrable functions
for given $x_c\in \check{D}_w$,
 this transformed equation is considered to be a {\it regular} Sturm-Liouville
 equation~\cite{Niessen,Zettl}, and hence $\psi$ and $P\psi'$ are continuous including
 the point $x=x_c$. Moreover, 
 we note from \eqref{SNM} that the small and large neutral modes,
 respectively, satisfy $\psi_+(x_c)=0$ and $(P\psi_-')(x_c)=0$
at the critical layer $x_c\in\check{D}_w$.
Base on this fact, we divide the domain into $[-L,x_c]$ and
$[x_c,L]$, and seek the small and large solutions on each side as follows.
By introducing the operator,
\begin{align}
 \Ope{E}(x_c,\lambda)\psi:=(|P|\psi')'-(\lambda+Q)|P|\psi,
\end{align}
that depends on the parameter $\lambda=k^2>0$, we solve 
\begin{align}
 \Ope{E}(x_c,\lambda_{L+})\psi=0,&\ \psi(-L)=0,\ \psi(x_c)=0,\label{L+}\\
 \Ope{E}(x_c,\lambda_{L-})\psi=0,&\ \psi(-L)=0,\ (|P|\psi')(x_c)=0,
\end{align}
on $[-L,x_c]$
and
\begin{align}
 \Ope{E}(x_c,\lambda_{R+})\psi=0,&\ \psi(L)=0,\ \psi(x_c)=0,\\
 \Ope{E}(x_c,\lambda_{R-})\psi=0,&\ \psi(L)=0,\ (|P|\psi')(x_c)=0,\label{R-}
\end{align}
on $[x_c,L]$. Because each of \eqref{L+}-\eqref{R-} is a regular Sturm-Liouville
problem with either Dirichlet or Neumann boundary condition at
$x=x_c$, 
Sturm's oscillation theorem~\cite{Zettl} guarantees that the
eigenvalues can be ordered as follows.
\begin{align*}
 \infty>\lambda_{L-}^{(1)}>\lambda_{L+}^{(1)}>\lambda_{L-}^{(2)}>\lambda_{L+}^{(2)}>\lambda_{L-}^{(3)}>\dots\rightarrow-\infty,\\
 \infty>\lambda_{R-}^{(1)}>\lambda_{R+}^{(1)}>\lambda_{R-}^{(2)}>\lambda_{R+}^{(2)}>\lambda_{R-}^{(3)}>\dots\rightarrow-\infty,
\end{align*}
where the superscript $(n)$ indicates the $n$th largest
eigenvalue. Here, only a finite number of positive eigenvalues
 is of our interest (since $\lambda=k^2>0$) and these eigenvalues depend on $x_c\in\check{D}_w$ continuously.
Again at the endpoints $x_c\in D_w\backslash\check{D}_w$ (namely, $J_R(x_c)=1/4$), the
degeneracy, $\lambda_{L-}^{(n)}=\lambda_{L+}^{(n)}$ and
$\lambda_{R-}^{(n)}=\lambda_{R+}^{(n)}$ ($n=1,2,\dots$), must occur.
 If the matching condition $\lambda_{L-}=\lambda_{R-}>0$
 between the left and right domains is satisfied for a specific
$x_c\in D_w$, there is a large neutral mode $\psi_-$ for such $c=U(x_c)$ and
$k=\sqrt{\lambda_{L-}}=\sqrt{\lambda_{R-}}$. Similarly, there exists a small neutral
mode $\psi_+$ if $\lambda_{L+}=\lambda_{R+}>0$ occurs for a specific $x_c\in D_w$.

\begin{figure}
\begin{center}
\includegraphics[width=60mm]{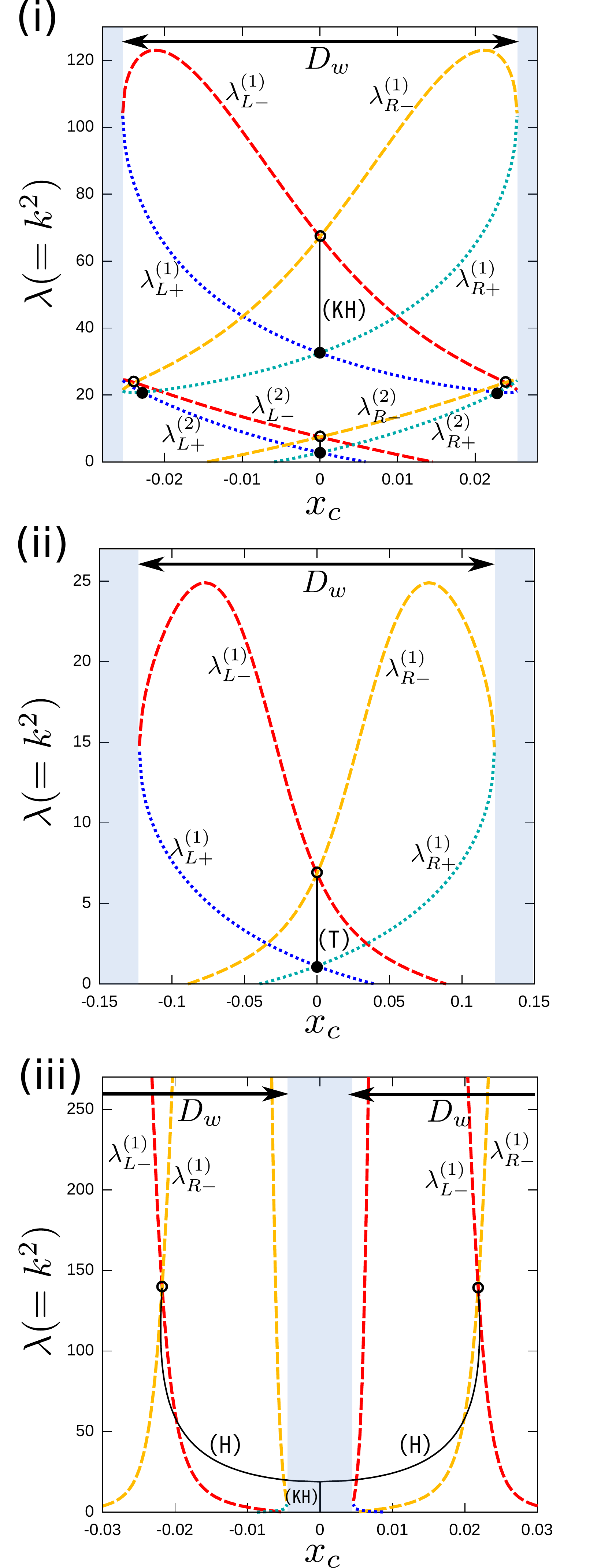} 
\end{center}
\caption{Numerically calculated $\lambda_{L\pm}^{(n)}$ and
 $\lambda_{R\pm}^{(n)}$, $n=1,2,\dots$, as functions of $x_c\in D_w$ for
 the examples of unstable flows (i)-(iii).} \label{unstable_flows}
\end{figure}

We have applied this approach numerically to the following three examples:
\begin{quote}
(i) $U=\tanh(10x)$, $N_B^2=22$,

(ii) $U=x$, $N_B^2=0.1+10x^2$,

(iii) $U=\tanh(10x)$, $N_B^2=30/(1+10^4x^2)$,
\end{quote}
on the common domain $[-L,L]=[-1,1]$. The results are shown in
Fig.~\ref{unstable_flows}(i)-(iii), where the shaded areas indicate the
exterior of $D_w$.
Several large and small neutral modes
are found, respectively, as indicated by the open and filled circle symbols in Fig.~\ref{unstable_flows}. We have
confirmed that unstable
modes indeed exist between these points along the black solid lines. The flows (i)-(iii)
are respectively unstable to
the Kelvin-Helmholtz (KH), Taylor (T) and Holmboe (H) instabilities as
indicated in Fig.~\ref{unstable_flows}. For the case (i), the
stratification induces three more unstable modes in addition to the KH instability
that originally exists in the mixing layer. As exemplified by these
modes  as well as the Taylor instability (ii), increasing the stratification (or $N_B^2$) generally tends to spawn
more unstable modes, whose eigenfunctions are more
oscillatory (i.e., $\lambda_{L\pm}^{(n)}$ and
 $\lambda_{R\pm}^{(n)}$ with larger $n$ show up on the positive
 side). 
  At the same time, however, the
 width of $D_w$ tends to be narrowed by increasing $N_B^2$, which
 in turn contributes
 to stabilization. In the case (iii), 
 $D_w$ is divided into two parts by the sharply peaked profile of $N_B^2$ (or density
 jump) and the KH mode is split into the Holmboe modes. Here, it should
 be noted that the unstable modes may exist outside $D_w$ while the
 neutral modes are forced to exist inside $D_w$.

\section{Sufficient Conditions for Stability}

Now, we recall that Howard~\cite{Howard} has obtained the upper bound of
${\rm Im}\,C$ as ${\rm Im}\,C<k^{-1}\max_{[-L,L]}\sqrt{U'^2/4-N_B^2}$,
which implies ${\rm Im}\,C\rightarrow0$ as
$k\rightarrow\infty$. On the other hand, we have shown the absence of neutral modes for
sufficiently large wavenumber $k$ satisfying either
$k^2>\lambda_{L-}^{(1)}$ or $k^2>\lambda_{R-}^{(1)}$ for all $x_c\in D_w$. Therefore, no unstable mode can exist for such large $k$.
Since the largest eigenvalues, $\lambda_{L-}^{(1)}$ and
$\lambda_{R-}^{(1)}$, of the Sturm-Liouville problem are
easily
calculated by the variational method, we can derive a stability condition in the
following form.
\begin{criterion}
The flow \eqref{assumption} is spectrally stable for the wavenumber $k$ that
 satisfies either
\begin{align}
k^2>-\min_{\psi(-L)=0}\frac{\int_{-L}^{x_c}|P|(\psi'^2+Q\psi^2)dx}{\int_{-L}^{x_c}|P|\psi^2dx},\label{va}\\
\mbox{ or }\quad
k^2>-\min_{\psi(L)=0}\frac{\int_{x_c}^L|P|(\psi'^2+Q\psi^2)dx}{\int_{x_c}^L|P|\psi^2dx},\label{vb}
\end{align}
for all $x_c\in D_w$.
\end{criterion}
Obviously, the flow is spectrally stable for all $k$ if the right hand sides of \eqref{va} and \eqref{vb} are
negative
for all $x_c\in D_{w1}$ and for all $x_c\in D_{w2}$, respectively, and
$D_w=D_{w1}\cup D_{w2}$
 (namely, $\lambda_{L-}^{(1)}$ and
$\lambda_{R-}^{(1)}$ are negative on $D_{w1}$ and $D_{w2}$, respectively).
Note that either $D_{w1}$ or $D_{w2}$ may be the null set and then \eqref{va} or
\eqref{vb} will be not applicable.

If $Q(x,x_c)\ge0$ both on $[-L,x_c]$ for all
$x_c\in D_{w1}$ and on $[x_c,L]$ for all
$x_c\in D_{w2}$, Criterion 1 immediately proves stability.
This observation enables us to derive
 further stability conditions as follows.

Let $a\in[-L,L]$ be the rightmost point in $D_{w1}$, i.e., $a=\max D_{w1}$.
By using the inequality
\begin{align}
 \frac{J_R(x_c)-J_R(x)}{U(x_c)-U(x)}\ge\min_{[-L,a]}\frac{J_R'}{U'}
\quad\mbox{for }x\le x_c\le a,
\end{align}
 we get
\begin{align}
  Q(x,x_c)\ge&\frac{U'^2}{c-U}\left[
-\left(\frac{1}{2}+\nu_c\right)\frac{U''}{U'^2}+\min_{[-L,a]}\frac{J_R'}{U'}
\right],
\end{align}
on $[-L,x_c]\subset[-L,a]$. 
The requirement $Q(x,x_c)\ge0$ on $[-L,x_c]$ for all
$x_c\in D_{w1}$ (for which $0\le\nu_c<1/2$) leads to \eqref{ma} below.
The similar inequality can be also derived
for $[x_c,L]\subset[b,L]$ in terms of the leftmost point $b$ in
$D_{w2}$. Thus, we obtain the following stability criterion:
\begin{criterion}
The flow \eqref{assumption} is spectrally stable if there exist
 $a,b\in[-L,L]$ such that
\begin{align}
 \max_{[-L,a]\atop\sigma=1/2,1}\left(\sigma\frac{U''}{U'^2}\right)\le\min_{[-L,a]}\left(\frac{J_R'}{U'}\right),\label{ma}\\
 \min_{[b,L]\atop\sigma=1/2,1}\left(\sigma\frac{U''}{U'^2}\right)\ge\max_{[b,L]}\left(\frac{J_R'}{U'}\right),\label{mb}
\end{align}
and $D_w\subset [-L,a]\cup[b,L]$ hold, where \eqref{ma} or \eqref{mb} may be omitted
when $a=-L$ or $b=L$, respectively.
\end{criterion}
We remark that, in \eqref{ma}, for example, we should adopt $\sigma=1/2$
 if $\max_{[-L,a]}U''<0$, or $\sigma=1$ otherwise.

From Criterion 2, it is straightforward to obtain the next criterion:
\begin{criterion}
 The flow \eqref{assumption} is spectrally stable if there exist
 $a,b\in[-L,L]$ such that
\begin{align}
\mbox{$U''\le0$ and $J_R'\ge0$ on $[-L,a]$,} \label{a}\\
 \mbox{$U''\ge0$ and $J_R'\le0$ on $[b, L]$,} \label{b}
\end{align}
and $D_w \subset [-L,a]\cup[b,L]$ hold, where \eqref{a} or \eqref{b} may be omitted
when $a=-L$ or $b=L$, respectively.
\end{criterion}
Recall that {\it unstratified} shear flows with no inflection point
($U''\ne0$) were shown to be stable by Rayleigh~\cite{Rayleigh}, and even
when a monotonic shear flow ($U'>0$) has one inflection point $U''(x_I)=0$, it
is still
stable if $U''<0$ on $[-L,x_I]$ and $U''>0$ on $[x_I,L]$  according to Fj\o rtoft~\cite{Fjortoft}.
Criterion 3 asserts that these
flows remain stable even with the effect of stratification, if the
additional condition on the sign of $J_R'$ is satisfied (where $a=b=x_I$ is
chosen for the Fj\o rtoft case).

Since $J_R'\lessgtr0$ is equivalent to $(N_B^2)'/N_B^2=\rho''/\rho'\lessgtr U''/U'$
by definition, the case of $N_B^2\equiv$ const. (or $\rho''\equiv0$) automatically
satisfies the condition on $J_R'$ of Criterion 3.
On the contrary, if $J_R$ takes a minimum value in the middle of
$[-L,L]$ and it is below $1/4$, Criterion 3 is violated and such a
stratification can destabilize an otherwise stable shear
flow, like the Taylor instability of Fig.~\ref{unstable_flows}(ii).
Criterion 3 is also violated by a $U'$ that has a local maximum in the middle
of $D_w$ (i.e., a local maximum of vorticity), which can cause 
KH instability like the example of Fig.~\ref{unstable_flows}(i).

\begin{figure}
\begin{center}
\includegraphics[width=60mm]{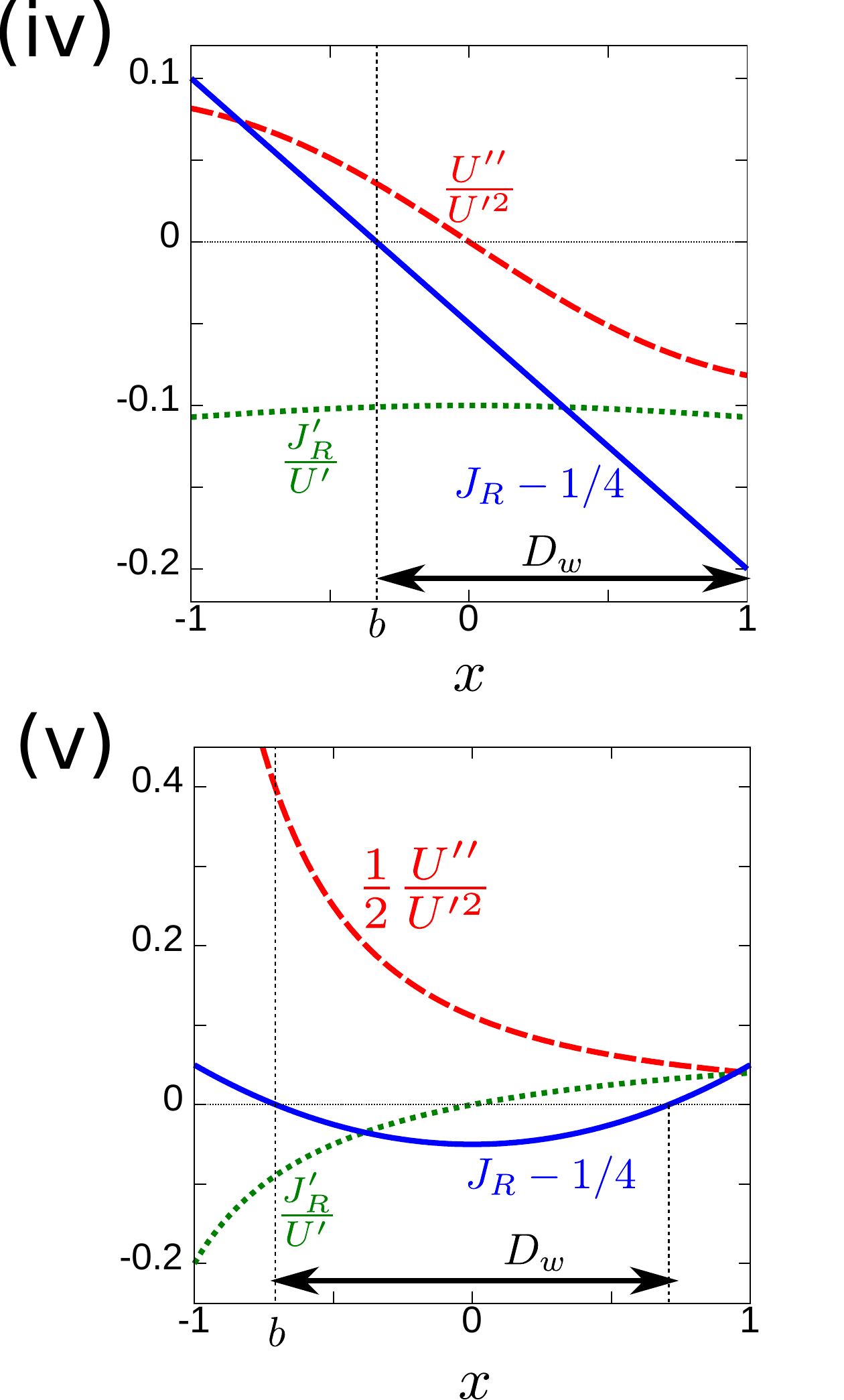} 
\end{center}
\caption{Examples of stable flows (iv) and (v) found by Criterion 2} \label{stable_flows}
\end{figure}

The more detailed Criterion 2 tells us that this destabilization
effect of $U''$ (or $J_R'$) may be suppressed by the stabilization
effect of $J_R'$ (or $U''$). For example, the flows,
\begin{quote}
(iv) $U=\arctan(x/2)+x$, $J_R=(1-3x/4)/5$,

(v) $U=(x+3/2)^2$, $J_R=(x^2+2)/10$,
\end{quote}
on $[-L,L]=[-1,1]$ are found to be stable by Criterion 2.
For the case (iv), the sign of $U''$ violates the Rayleigh-Fj\o rtoft
criterion and $U'$ has a maximum in $D_w$, but the stratification $J_R$ contributes to stabilization so
that the inequality \eqref{mb} with $D_w=[b,L]$ holds, as shown in
Fig.~\ref{stable_flows}. Conversely, for the case (v), $J_R$ has a minimum
that is below $1/4$, but $U''>0$ is large enough to satisfy \eqref{mb}
with $D_w\subset [b,L]$.

\section{Concluding Remarks}

\begin{figure}
\begin{center}
\includegraphics[width=70mm]{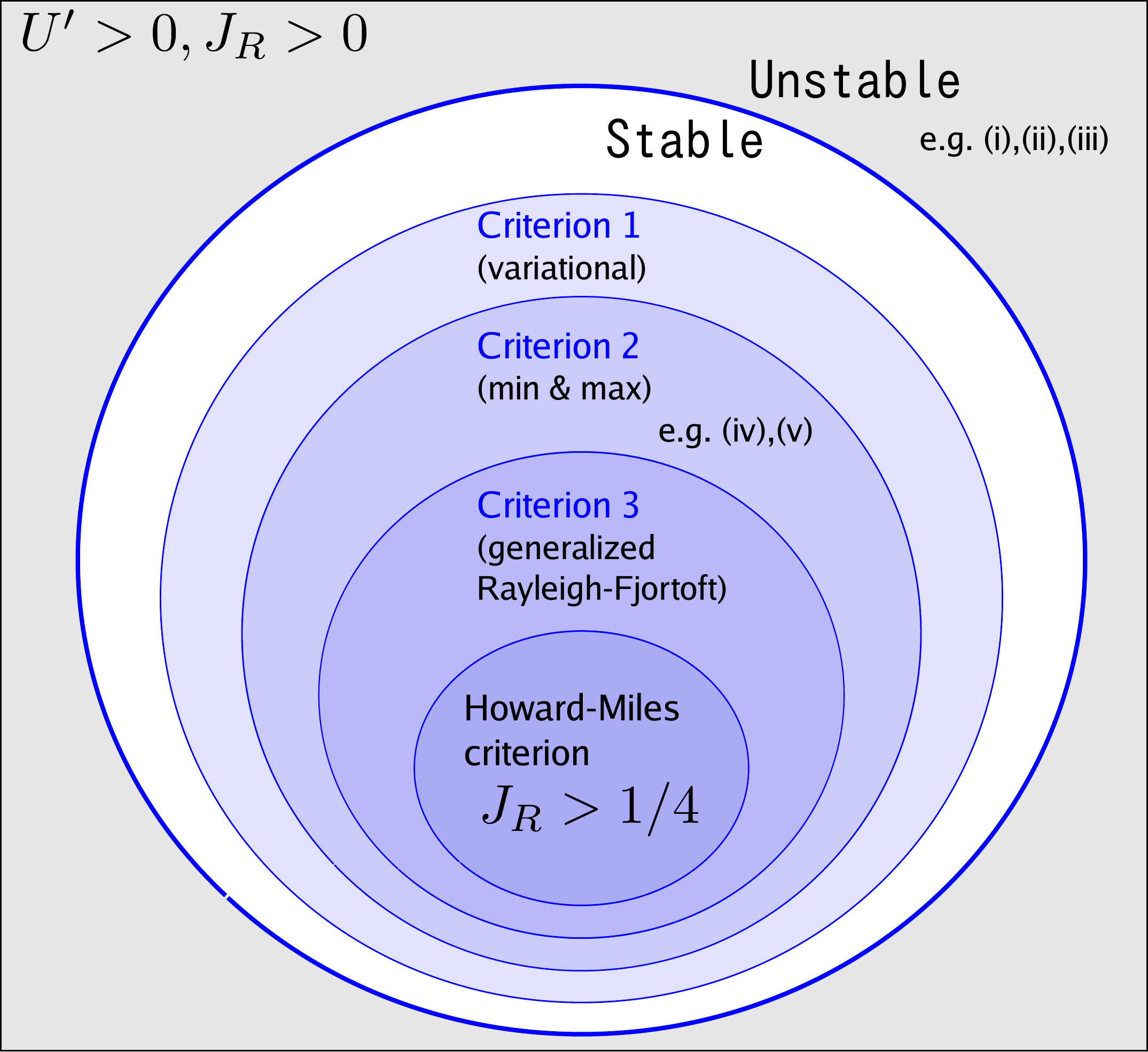} 
\end{center}
\caption{Venn diagram of sufficient stability criteria} \label{venn_diagram}
\end{figure}

In summary, by using the method developed in this letter, we can
efficiently find singular neutral modes that are
embedded in the continuous spectrum. The unstable
ranges of wavenumber are searched for by
solving the regular Sturm-Liouville problems \eqref{L+}-\eqref{R-} for each $x_c\in D_w$,
which requires less labor than solving the TG equation (i.e., a
non-self-adjoint eigenvalue problem) for each $k>0$ directly.
With the help of numerical computation, this method will be useful for
determining stability boundaries for various flows and understanding
their instability mechanisms.
 Based on this method, we have
derived new stability criteria (Criteria 1,2, and 3) as extensions of the
Rayleigh-Fj\o rtoft criterion to stratified flows, which are of course
improvements of the Howard-Miles criterion
since the local Richardson number is allowed to be less
than 1/4 (see Fig.~\ref{venn_diagram}).

We expect that the present
approach could yields stability boundaries of non-monotonic shear flows
and locally unstratified (or constant) densities
with additional or different treatments for the singularity.
A few specific examples were
investigated in earlier works~\cite{Huppert,Churilov,Rees}, which in particular
 suggests that the singularity at the critical layer should be replaced
by that of Rayleigh's equation when $N_B^2(x_c)=0$.
By making use of the abundant knowledge about Rayleigh's equation~\cite{Hirota},
more detailed
stability criteria
tailored for these applications will be discussed elsewhere.

This work was supported by JSPS KAKENHI Grant Number~25800308 and by JSPS Strategic Young Researcher Overseas
Visits Program for Accelerating Brain Circulation \# 55053270. PJM was
supported by DOE
Office of Fusion Energy Sciences, under DE-FG02-04ER-54742.

\end{document}